\documentclass[twocolumn,prl,superscriptaddress]{revtex4-1}
\usepackage{graphicx}
\usepackage{amssymb}
\usepackage{amsmath}
\usepackage{epsfig}
\usepackage{color}
\usepackage{natbib}
\usepackage{mathtools}
\usepackage[colorlinks,linkcolor=blue,anchorcolor=blue,citecolor=blue,urlcolor=blue]{hyperref}
\usepackage[left]{lineno}
\usepackage{blindtext}
\usepackage{color, soul}
\usepackage{setspace}
\usepackage{fancyhdr}

\begin{document}

\title{Experimental implementation of universal holonomic quantum computation on solid-state spins with optimal control}
\author{Yang Dong}
\author{Shao-Chun Zhang}
\author{Yu Zheng}
\author{Hao-Bin Lin}
\author{Long-Kun Shan}
\author{Xiang-Dong Chen}
\affiliation{{CAS Key Laboratory of Quantum Information, University of Science and Technology of China, Hefei, 230026, P.R. China}}
\affiliation{{CAS Center for Excellence in Quantum Information and Quantum Physics, University of Science and Technology of China, Hefei, 230026, P.R. China}}
\author{Wei Zhu}
\author{Guan-Zhong Wang}
\affiliation{{Hefei National Laboratory for Physical Science at Microscale, and Department of Physics, University of Science and Technology of China, Hefei, Anhui 230026, P. R. China}}\author{Guang-Can Guo}
\author{Fang-Wen Sun}
\email{fwsun@ustc.edu.cn}
\affiliation{{CAS Key Laboratory of Quantum Information, University of Science and Technology of China, Hefei, 230026, P.R. China}}
\affiliation{{CAS Center for Excellence in Quantum Information and Quantum Physics, University of Science and Technology of China, Hefei, 230026, P.R. China}}

\begin{abstract}
Experimental realization of a universal set of quantum logic gates with high-fidelity is critical to quantum information processing, which is always challenging by inevitable interaction between the quantum system and environment. Geometric quantum computation is noise immune, and thus offers a robust way to enhance the control fidelity. Here, we experimentally implement the recently proposed extensible nonadiabatic holonomic quantum computation with solid spins in diamond at room-temperature, which maintains both flexibility and resilience against decoherence and system control errors. Compared with previous geometric method, the fidelities of a universal set of holonomic single-qubit and two-qubit quantum logic gates are improved in experiment. Therefore, this work makes an important step towards fault-tolerant scalable geometric quantum computation in realistic systems.
\end{abstract}
\maketitle

Based on quantum superposition and entanglement \cite{suter2016colloquium,waldherr2014quantum,streltsov2017colloquium,dong2018non}, quantum information processing can offer a dramatic speed-up over classical method in simulations \cite{van2012decoherence,wu2019observation}, prime factoring \cite{peng2008quantum}, database searching \cite{zhang2020efficient}, machine learning \cite{carleo2019machine}, and so on. Currently, as quantum coherence \cite{streltsov2017colloquium} is fragile under noises from either realistic environment or control fields, one of the most urgent requirements of quantum computation is to realize error-resilient universal set of single-qubit and two-qubit gates \cite{rong2015experimental,hegde2020efficient,suter2016colloquium,casanova2016noise,zhang2014protected,zhang2015experimental,golter2014protecting}. Comparing with the dynamical control that depends on the instantaneous value of Hamiltonian of quantum system, geometric phases depend only on the global property of enclosed path, and thus have built-in noise-resilient features against certain local noises \cite{sjoqvist2012non,zu2014experimental,xu2018single,sekiguchi2017optical,kleissler2018universal}. Moreover, if the quantum system has degenerate eigenstates, the geometric phase can act on the degenerate subspace \cite{duan2001geometric,liu2017superadiabatic} and is a key recipe for achieving high-fidelity quantum gates and constructing full geometric quantum computation (GQC).

Originally, GQC is constructed by adiabatic evolutions to avoid transitions between different energy level states and the logic gates only depend on the ratio of control parameters, which are immune to the fluctuation of their absolute values \cite{huang2019experimental,guery2019shortcuts}. However, the adiabatic condition simultaneously requires long run time, and thus decoherence will introduce considerable errors \cite{yan2019experimental,xu2020experimental,guery2019shortcuts}. To overcome such an intrinsic disadvantage, nonadiabatic holonomic quantum computation (NHQC) schemes are proposed in theory by driving system Hamiltonian with time-independent eigenstates \cite{sjoqvist2012non}. Compared to the adiabatic case, this new scheme is fast and has been experimentally demonstrated in superconducting circuits \cite{abdumalikov2013experimental,yan2019experimental}, nuclear magnetic resonance \cite{feng2013experimental,zhu2019single}, solid spin system \cite{zu2014experimental,nagata2018universal,arroyo2014room,zhou2017holonomic,yale2016optical}, and trapped ion \cite{leroux2018non,toyoda2013realization}. However, because of the challenges of exquisite control among quantum systems, the systematic errors will introduce additional fluctuating phase shifts and lead to a serious reduction of fidelity than dynamical gates \cite{johansson2012robustness,zheng2016comparison,huang2019experimental}. Fortunately, an extensible version--NHQC+ scheme, which is robust against the control errors and compatible with common optimization techniques for improving the logic gates fidelity, was recently proposed \cite{liu2019plug,li2020fast}. And in experiment, only single qubit operation was demonstrated in a superconducting circuit \cite{yan2019experimental} and trapped ion \cite{ai2020experimental}. It is still a non-trivial task to realize two-qubit gates based on the NHQC+ scheme due to the challenging requirement of exquisite control of complicated energy level structure \cite{liu2019plug,li2020fast,yan2019experimental,ai2020experimental,xu2020experimental}.

In this Letter, based on the NHQC+ scheme, we experimentally demonstrate high fidelity single-qubit and two-qubit geometric quantum gates, which together make a universal gate set, all by nonadiabatically manipulating spin states in a diamond nitrogen-vacancy (NV) center using a significantly simplified single-loop evolution. By shaping both amplitudes and phases of two microwave fields, we show explicitly in experiments that the NHQC+ scheme is robust against significant variations in control parameters. As a typical hybrid solid spin system, the NV center electronic spin ($S{\text{ = }}1$) coupling with a nitrogen nuclear spin ($I{\text{ = }}1$) shows several intrinsic properties for the NHQC+ scheme. It also poses some challenges in experiment, mostly because the characteristic properties of the two types of spins differ by $3$ orders of magnitude. The electron spin dephasing time is usually shorter than typical durations of gates applied to the nitrogen spin. Here, we apply dynamical decoupling method \cite{liu2019plug,dong2016reviving} in the scheme to achieve high fidelity non-trivial nonadiabatic holonomic two-qubit gates in experiment. Therefore, based on the NHQC+ scheme, a programmable solid-state quantum information processor can be constructed efficiently at room-temperature. Moreover, the geometric phase has a close relationship with topological phase. The demonstration of the robust holonomic quantum gates is a critical step towards the realization of topological quantum computation \cite{nayak2008non,li2019interfacing}.

\begin{figure}[tbp]
\centering
\textsf{\includegraphics[width=8.7cm]{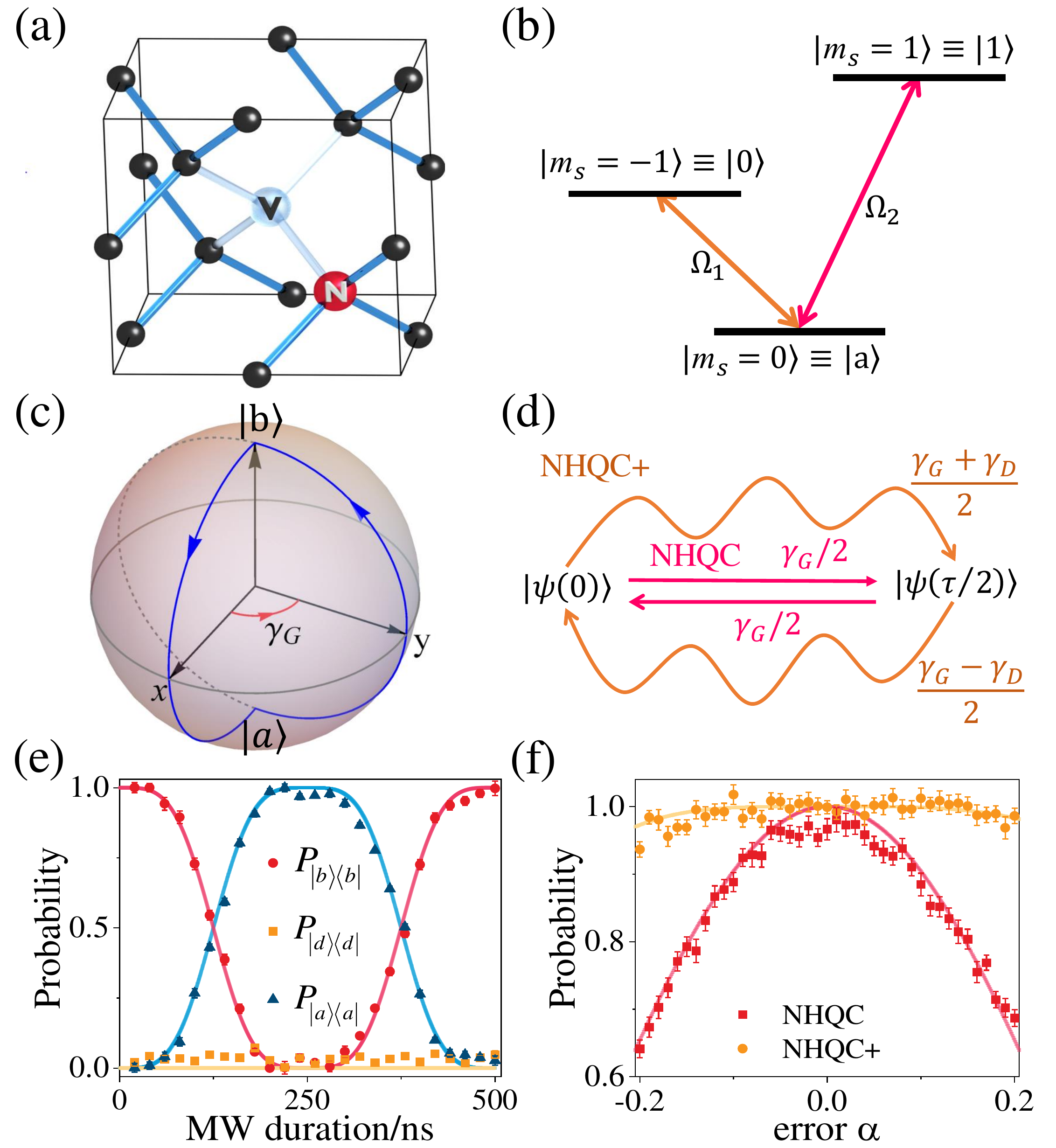}}
\caption{(a) Schematic of the unit cell of diamond including an NV center. (b)
Encoding of a qubit in the spin-triplet ground state and the microwave coupling configuration. (c) Conceptual explanation for geometric quantum operation in Bloch sphere, showing the geometric phase which equals half of the enclosed solid angle ${\gamma _G}$. (d) Schematic of different paths for the NHQC (red) and NHQC+ (brown) with optimization methods. The evolution is divided into two steps: first evolving from bright state $\left| b \right\rangle $ to auxiliary state $\left| a \right\rangle $ and then back with an accumulation of a pure geometric phase. (e) A holonomic $X$ gate on a bright state.
Experimental results (dot) are fitted well with numerical
simulations (solid lines) by setting $\eta  =1$. (f) Performance of an $X$ gate with control errors ($\alpha $) for NHQC and NHQC+ schemes with initial state $\left| 0 \right\rangle $. The envelopes
of the two driving fields of NHQC are truncated Gaussian pulses \cite{zu2014experimental,xu2018single}. }
\label{fig1}
\end{figure}

At first, we explain how to implement the holonomic single-qubit gates in experiment. We manipulate the electron spin states of NV center in a synthetic diamond \cite{zhao2020improving,dong2020quantifying} at room temperature as shown in Fig. \ref{fig1} (a).
We encode $\left| {{m_s} =  - 1} \right\rangle  \equiv \left| 0 \right\rangle $ and
$\left| {{m_s} =  1} \right\rangle  \equiv \left| 1 \right\rangle $ as the qubit basis states, and use $\left| {{m_s} = 0} \right\rangle  \equiv \left| a \right\rangle $ as an ancillary state for the geometric manipulation, as shown in Fig. \ref{fig1} (b). The electron spin state is initialized to $\left| {{m_s} = 0} \right\rangle $ by optical pumping and read out by identifying distinct fluorescence intensity of the states after $300$ ns green laser pulse. We apply a magnetic field of $378$ G along the NV axis using a permanent magnet. Under this magnetic field, the nearby nuclear spins can be optically polarized through level anticrossing, enhancing the coherence time of the electron spin \cite{dong2018non,li2018enhancing} with $T_2^{\text{*}} \approx 8$ $\mu s$ and  ${T_2} \approx {600}$ $\mu s$, which are obtained from Ramsey and spin-echo signal, respectively.

We apply microwave (MW) pulses to couple the transition from the ancillary level $\left| a \right\rangle $ to the qubit states $\left| 0 \right\rangle $ and $\left| 1 \right\rangle $ with Rabi frequencies ${\Omega _1}\left( t \right)$ and ${\Omega _2}\left( t \right)$ respectively, as shown in Fig. \ref{fig1} (b). The Hamiltonian for the couplings among these three levels
takes the form
\begin{equation}
\begin{aligned}
{H_1} &= \left[ {\frac{{{\Omega _1}(t)}}{2}{e^{ - i{\phi _1}}}|0\rangle {\text{ + }}\frac{{{\Omega _2}(t)}}{2}{e^{ - i{\phi _2}}}|1\rangle } \right]\left\langle {\text{a}} \right| + {H}.{c.}\\
            &= {}\frac{\Omega }{2}{e^{ - i{\phi _1}}}\left| {\text{b}} \right\rangle \langle 0|{+ H}.{c.} \text{,}\\
\end{aligned}
\end{equation}
where $\Omega {  =  }\sqrt {\Omega _1^2 + \Omega _{2}^2} $, and the bright state is $\left| {b} \right\rangle  = \sin \frac{\theta }{2}\left| 0 \right\rangle  - \cos\frac{\theta }{2}{{e}^{i\phi }}\left| 1 \right\rangle $ with $\tan\frac{\theta }{2} = \frac{{{\Omega _1}}}{{{\Omega _2}}}$ and $\phi  = {\phi _1} - {\phi _2} - \pi $. In the NHQC+ scheme, $\theta$ and $\phi $ are time independent. Correspondingly, the dark state $\left| d \right\rangle  = \cos\frac{\theta }{2}\left| 0 \right\rangle  + \sin \frac{\theta }{2}{{\text{e}}^{ i\phi }}\left| 1 \right\rangle $ is decoupled from the
$\left\{ {\left| b \right\rangle ,\left| a \right\rangle } \right\}$ subspace.
Here, we choose the dark state and an orthogonal state $|\psi \rangle  = {e^{ - if/2}}\left( {\cos\frac{\beta }{2}{e^{ - i\varphi /2}}\left| b \right\rangle  + \sin\frac{\beta }{2}{e^{i\varphi /2}}\left| a \right\rangle } \right)$ to demonstrate how to build up universal arbitrary holonomic single-qubit gates. In the NHQC+ scheme, any complete set of basis vectors satisfies the following conditions: ${\rm i})$ the cyclic evolution path ${\Pi _1}\left( 0 \right) = {\Pi _1}\left( \tau  \right) = \left| d \right\rangle \left\langle d \right|$, and ${\Pi _2}\left( 0 \right) = {\Pi _2}\left( \tau  \right) = \left| b \right\rangle \left\langle b \right|$; ${\rm ii})$  the von Neumann equation $\frac{d}{{dt}}{\Pi _k}\left( t \right) =  - i\left[ {{H_1},{\Pi _k}} \right]$, where ${\Pi _1}(t) = \left| d \right\rangle \left\langle d \right|$ and ${\Pi _2}(t) = \left| \psi  \right\rangle \left\langle \psi  \right|$ denote the projectors. Using the von Neumann equation, we obtain the following
coupled differential equations \cite{ai2020experimental,li2020fast}:
$\dot f = \frac{{\dot \varphi }}{{\cos\beta }}$, $\dot \beta  = \Omega \sin\left( {\varphi  + {\phi _1}} \right)$, and $\dot \varphi  = \Omega \cot\beta \cos\left( {\varphi  + {\phi _1}} \right)$,
where the dot represents time differential.

As shown in Fig. \ref{fig1} (c), we adopt a single-loop evolution path to induce a pure geometric phase. Specifically, during the whole cyclic evolution time of $\tau$, we set $\beta=\pi \sin ^{2}(\pi t / \tau )$ and $f=\eta[2 \beta-\sin (2 \beta)]$. During the first time interval $[0,\tau /2)$, we set $\varphi ({t = 0}) = 0$. For the following part $t \in [\tau /2,\tau ]$, we set $\varphi \left( {t = \tau /2} \right) = {\gamma _G},$ and $\varphi(\tau)=\gamma_G$ with a constant angle $\gamma_G$. Moreover, we invert the dynamical phase of the latter interval to be opposite to the formal one and thus no dynamical phase will be accumulated finally. It is quite different with the original NHQC scheme as shown Fig. \ref{fig1} (d), where the dynamical phase at each moment is zero due to stringent requirements \cite{sjoqvist2012non,zu2014experimental,xu2018single,sekiguchi2017optical}. As a result, the corresponding evolution operator based on the NHQC+ scheme takes the form $U(\tau,0) = \left| d \right\rangle \left\langle d \right| + {e^{i\gamma_G }}\left| b \right\rangle \left\langle b \right|$ in the $\left\{ {\left| d \right\rangle ,\left| b \right\rangle } \right\}$ subspace, which is an arbitrary holonomic single-qubit gate in the computation basis $\left\{ {\left| 0 \right\rangle ,\left| 1 \right\rangle } \right\}$ as
\begin{equation}
U(\theta, \phi, \gamma)=e^{i \frac{\gamma_G}{2}} e^{-i \frac{\gamma_G}{2} \mathbf{n} \cdot \sigma} \text{.}
\end{equation}
It describes a rotation operation around the axis $\mathbf{n}=(\sin \theta \cos \phi, \sin \theta \sin \phi, \cos \theta)$ by an angle $\gamma_G$, with a global phase factor $\exp (i\gamma_G /2)$.

\begin{figure}[tbp]
\centering
\textsf{\includegraphics[width=8.7cm]{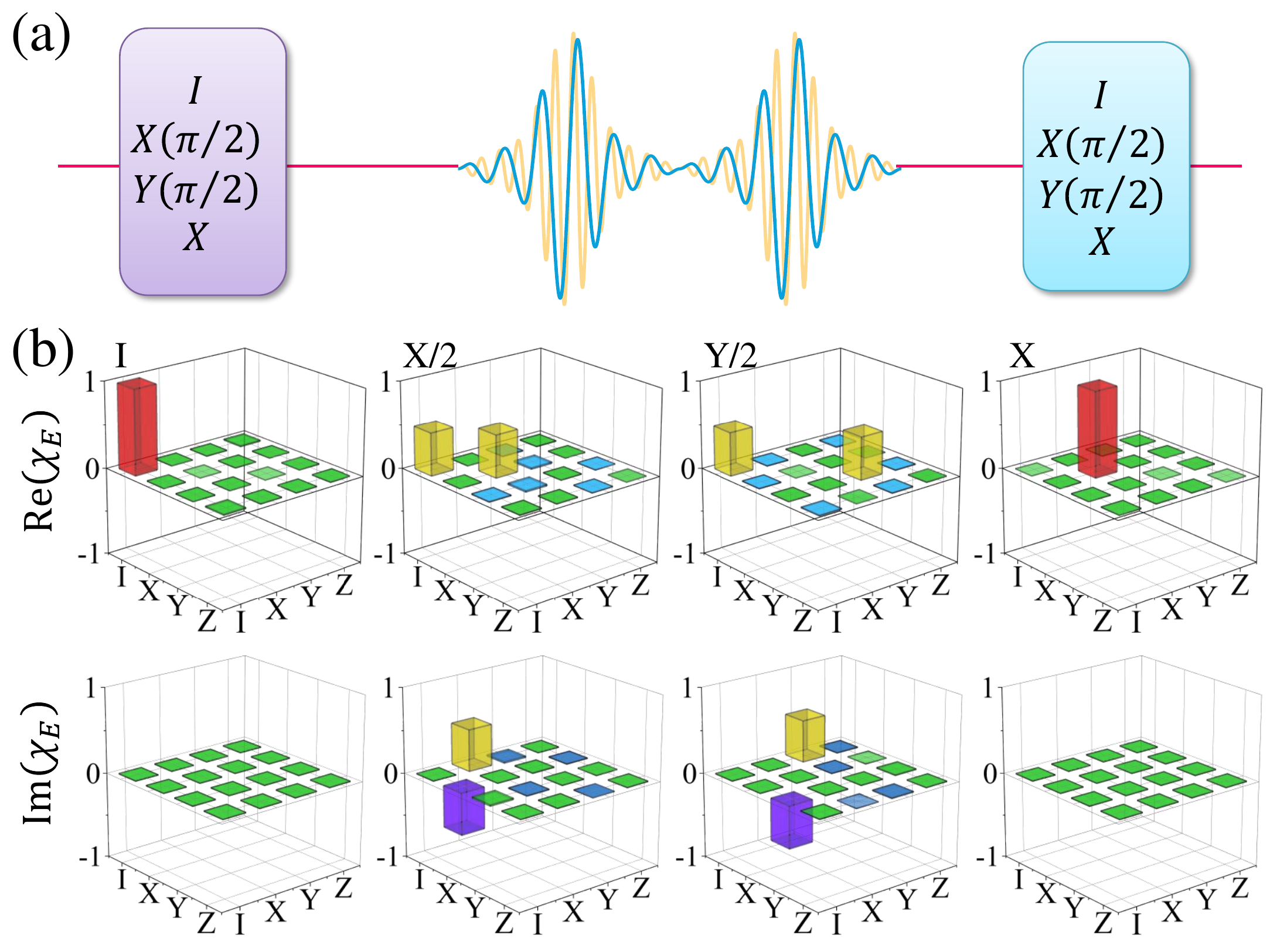}}
\caption{(a) The experimental pulse sequence to characterize the performance of the single-qubit gate. (b) QPT results of single-qubit gate. Real and imaginary parts of ${\chi _{\exp }}$ for specific gates $I$, $X/2$, $Y/2$, and $X$, giving an average process fidelity of $0.978(5)$. The labels in the $x$ and $y$ axes correspond to the operators in the basis set $\left\{ {I,{\sigma _x},{\sigma _y},{\sigma _z}} \right\}$ of $\left\{\left| 0 \right\rangle ,\left| 1 \right\rangle \right\}$ subspace. }
\label{fig2}
\end{figure}

In the experiments, we first observe the behavior of the states driven by ${H_1}$. We initialize the qubit to the bright state $\left| b \right\rangle  = \frac{{\left| 0 \right\rangle  - \left| 1 \right\rangle }}{{\sqrt 2 }}$ and set $\theta=\pi/2$, $\phi=0$, and $\gamma_G=\pi$ to construct an $X$ gate. Then the bright state evolves under the $X$ gate and is projected to $\left| a \right\rangle$, $\left| b \right\rangle $ and the dark state $\left| d \right\rangle  = \frac{{\left| 0 \right\rangle  + \left| 1 \right\rangle }}{{\sqrt 2 }}$. The experimental results are shown in Fig. \ref{fig1} (e).  And at end the of pulse sequence, the bright state would acquire a pure geometric phase. Also, the bright state is always decoupled with the dark state as ${H_1}\left| d \right\rangle  = 0$.
Furthermore, to show the robustness of the NHQC+ scheme, we perform an $X$-gate operation flipping $\left| 0 \right\rangle $ to $\left| 1 \right\rangle $ with an error in the amplitude of the control pulse, where the maximum Rabi frequency ${\Omega _{\max }}$ has some variation: ${\Omega _{\max }} = (1 + \alpha )\Omega $. The theoretical and experimental probabilities of the final state in $\left| 1 \right\rangle $ are plotted as a function of $\alpha $ in Fig. \ref{fig1} (f). Obviously, it clearly shows that the optimal NHQC+ is more robust than previous scheme for a broad range of pulse error \cite{ai2020experimental,xu2020experimental}.

\begin{figure}[tbp]
\centering
\textsf{\includegraphics[width=8.6cm]{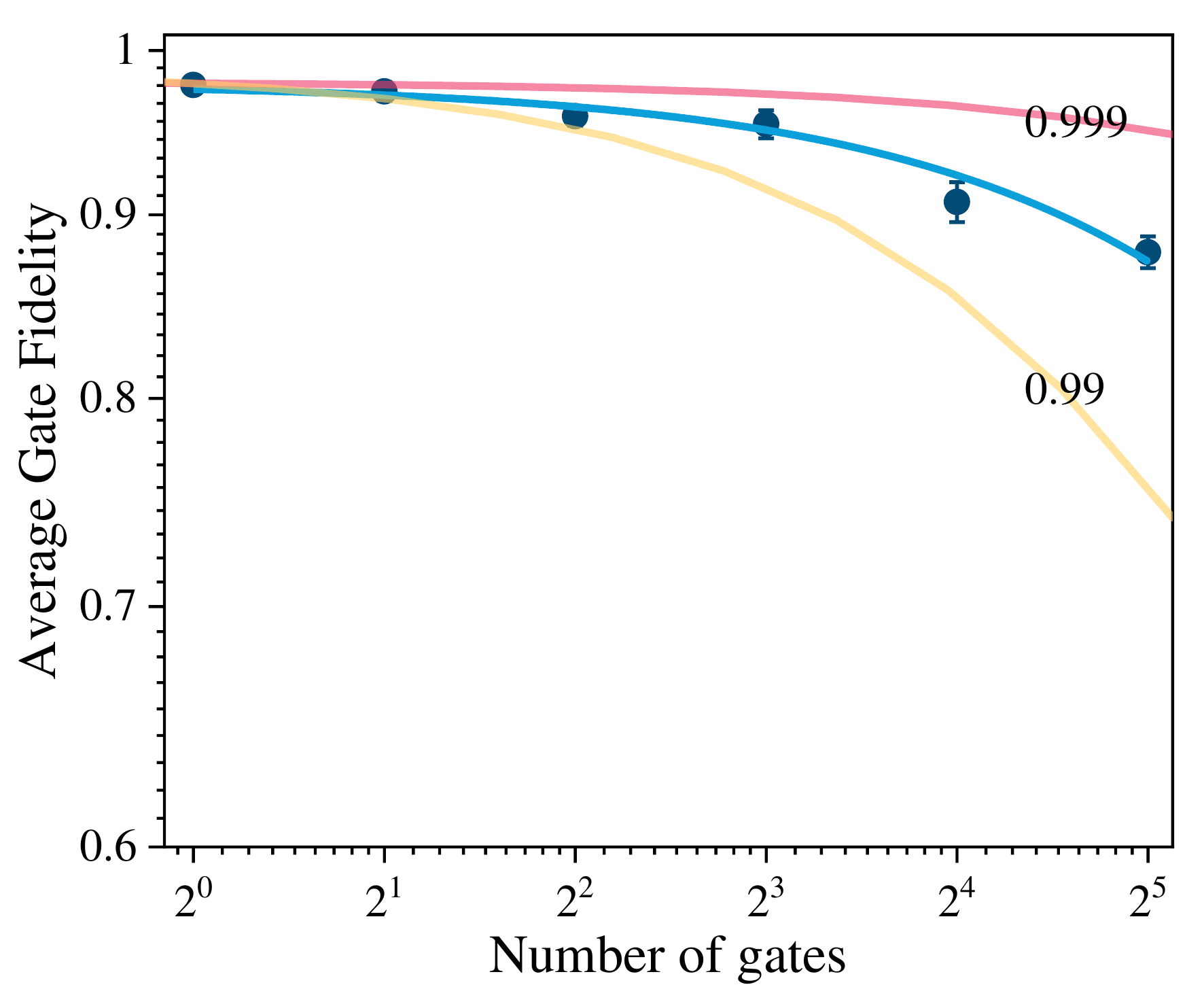}}
\caption{The decay of the fidelity of nonadiabatic holonomic $Y/2$ gates obtained via QPT. Experimental data (red dots) are fitted
by $F = \left[ {1 + \left( {1 - {\varepsilon _{if}}} \right){{\left( {1 -2p} \right)}^N}} \right]/2$ with solid curve and ${{\varepsilon _{if}}}$ describes
errors in state preparation and measurement. The other two curves show simulation results with $F_a=0.99$ and $F_a=0.999$, respectively.}
\label{fig3}
\end{figure}

\begin{figure*}[tbp]
\centering
\textsf{\includegraphics[width=15cm]{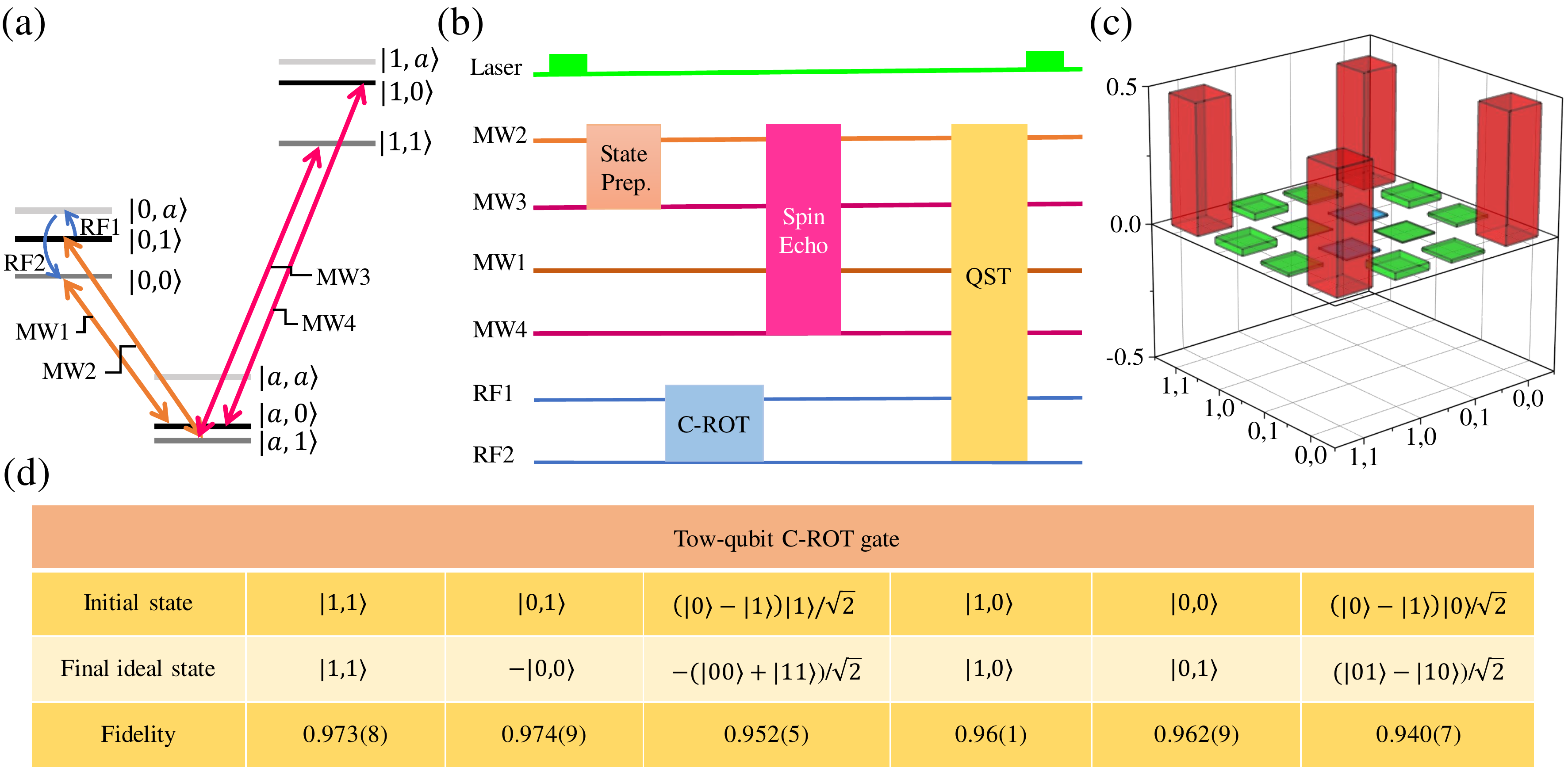}}
\caption{(a) The level structure of the electron and nuclear spins for the geometric CROT gate and the MW and RF coupling configuration.
(b) Time sequence for the implementation and verification of the geometric CROT gate between electron and nuclear spins. The CROT gate is implemented by applying with RF1 and RF2 pulses simultaneously. The duration time for CROT gate is 42 $\mu s$. MW pulses are used to implement a spin echo to increase the coherence time. To verify the CROT gate, we use a combination of MW1-MW4 and a RF to implement QST. (c) The matrix elements of the output density operator reconstructed through QST, where the geometric CROT is applied to the product state $\frac{{\left| 0 \right\rangle  - \left| 1 \right\rangle }}{{\sqrt 2 }}\left| 1 \right\rangle$. (d) Measured fidelities of the output states of the NHQC+ CROT gate with six typical input states.}
\label{fig4}
\end{figure*}

Then we demonstrate the implementation of the arbitrary holonomic gates in a single-loop path with NV center based on the similar procedure discussed above with $\eta  = 0.4$ \cite{SM}. We characterize the holonomic single-qubit gate through a standard quantum process tomography (QPT) method \cite{dong2020quantifying,ai2020experimental,SM} with experimental sequences shown in Fig. \ref{fig2} (a). The experimental process matrices ${\chi _{\exp }}$ of four specific geometric gates $I$, $X/2$, $Y/2$ and $X$ are shown in Fig. \ref{fig2} (b), with the fidelities of $0.983(4)$, $0.975(5)$, $0.978(5)$, and $0.977(5)$, respectively. Here, the process fidelity is calculated through $F = \left| {{Tr }\left( {{\chi _{\exp }}\chi _{id}^\dag } \right)} \right|$ by comparing with the idea operation $\chi _{id}$. The gates fidelities are better than previous results \cite{zu2014experimental}. As the original QPT is implemented by dynamical operation \cite{arroyo2014room} with long run time, it is more prone to noise, parameter imperfection and decoherence effects than holonomic gate. However, for the NHQC+ scheme in our experiment, the gate time is much smaller than the coherence time of NV center \cite{rong2015experimental}. Here, we can implement $Y/2$ gates successively and perform QPT when the gates are repeated as shown in Fig. \ref{fig3}. And the average fidelity \cite{zu2014experimental,rong2015experimental,rong2014implementation} of $N$ NHQC+ gates can be calculated as:
$F_N=\frac{1}{2}+\frac{1}{2}(1-2p)^{N}=\frac{1}{2}+\frac{1}{2} \exp \left(-T / T_{1 \rho}\right)$, where $T{\text{ = }}N\tau $ and $p$ is the average error per gate. We can estimate the average fidelity of each $Y/2$ gates to be ${F_a} = 1 - p = 0.9961(4)$ from the experimental data, which is close to the threshold of the fault-tolerant quantum computation \cite{rong2015experimental,rong2014implementation}. By lowing the temperature \cite{bar2013solid,astner2018solid}, ${T_{1,\rho }}$ and $T_{2}$ can be prolonged by $3-6$ orders of magnitude in proportion to ${T_{1}}$, which means the performance of the NHQC+ schema can be improved to touch the threshold of the fault-tolerant quantum computation.

To realize a universal quantum computation, a two-qubit gate is also necessary. Here, we use the non-zero host nitrogen nuclear spin \cite{li2018enhancing,geng2016experimental,lu2020observing} ($I{\text{ = }}1$ for $^{14}N$) of the NV center as the target qubit, as shown in Fig. \ref{fig4} (a). The Hamiltonian can be expressed as:
\begin{equation}
{H_s} = DS_z^2 + {\gamma _e}B{S_z} + PI_z^2 + {\gamma _n}B{I_z} + A_{zz}{S_z}{I_z} \text{.}
\end{equation}
Here, $S_{z}$ and $I_{z}$ are $z$ components of the spin-1 operators for the electron and $^{14}N$ nuclear spins, respectively. By applying state-selective MW and radio-frequency (RF) pulses, we can couple different energy levels. For arbitrary sublevels of electron spin ground state, the $^{14}N$ nuclear spin has a lambda ($\Lambda $) energy structure, which is preferable to realize the NHQC+ scheme \cite{liu2019plug,li2020fast}. Similarly, we encode $\left| {{m_I} =  - 1} \right\rangle  \equiv \left| 0 \right\rangle $ and $\left| {{m_I} =  1} \right\rangle  \equiv \left| 1 \right\rangle $ as the qubit basis states and use $\left| {{m_I} = 0} \right\rangle  \equiv \left| a \right\rangle $ as an ancillary state for the geometric manipulation of the $^{14}N$ nuclear spin.

For a two-qubit gate, the target qubit is changed depending on the state of the control qubit. The unitary operation for a controlled rotation (CROT) gate is equivalent to a controlled not (CNOT) gate \cite{jelezko2004observation,zhang2015experimental}. A typical CROT gate
can be represented as
\[{U_{\text{CROT}}} = \left( {\begin{array}{*{20}{c}}
  \mathbf{{{R_y}(\pi )}}&\mathbf{0} \\
  \mathbf{0}&\mathbf{1}
\end{array}} \right)\text{,}\]
where $\mathbf{1},\mathbf{0}$, and $\mathbf{{{R_y}(\pi )}}$ represent $2 \times 2$ matrices corresponding to the unit operator, zero matrix, and a rotation matrix around the $y$ axis with ${R_y}\left( \pi  \right) = {e^{ - i\pi {\sigma _y}/2}}$. Fig. \ref{fig4} (b) shows the circuit model of the CROT gate.
At first, we initialize the hybrid quantum system to the auxiliary state $ \left| {a,1} \right\rangle $ by the green laser. Then by employing MW operation, the bright state of electron spin $\left| {b,1} \right\rangle $ is prepared. Due to the magnetic dipole interaction ${A_{zz}}{S_z}{I_z}$, the energy level splitting of $^{14}N$ nuclear spin depends on electron spin state. Assisting with state-selective RF pulses, we can apply holonomic $Y$ gate on $^{14}N$ nuclear spin, which depends on the electron spin state, as shown in Fig. \ref{fig4} (a). Hence, for such a solid hybrid spin system, the two-qubit CROT gate can be implemented similarly to the single-qubit case. However, the duration of one CROT gate is long that decoherence effect destroys a significant part of the quantum information. To correct that, we apply dynamical decoupling method in the scheme with spin echo pulses \cite{zhang2014protected,zhang2015experimental,zu2014experimental,huang2019experimental} at the middle of the whole circuit with the time sequence shown in Fig. \ref{fig4} (b). As an example, with the initial state $\frac{{\left| 0 \right\rangle  - \left| 1 \right\rangle }}{{\sqrt 2 }}\left| 1 \right\rangle$, the final state is detected by quantum state tomography (QST). The experimental results are shown in Fig. \ref{fig4} (c). Compared with the idea final state $\frac{{\left| {00} \right\rangle {\text{ + }}\left| {11} \right\rangle }}{{\sqrt 2 }}$, the measured entanglement fidelity is $95.2(5)\%$, which unambiguously confirms the entanglement and is higher than the previous protocols with same system \cite{zu2014experimental,huang2019experimental}. Furthermore, the measured final state fidelities are listed in Fig.\ref{fig4} (d) with six typical input states for the holonomic two-qubit CROT gate.

In conclusion, we have experimentally demonstrated a universal set of robust holonomic gates using individual spins, which paves the way for all-geometric quantum computation with a solid-state system. The noise-resilient feature of the realized single-qubit geometric gates is also verified by comparing the performances of NHQC and NHQC+ schemes. The distinct advantages of the realized NHQC+ scheme indicate they are promising candidates for robust quantum information processing. Moreover, a nontrivial high-fidelity two-qubit CROT gate is preformed with this geometric protocol enhanced by dynamical decoupling method.
Since the electron and nuclear spins of different NV centers can be wired up quantum mechanically and magnetically by direct dipole interaction \cite{dolde2013room,bradley2019ten}, phonon-induced spin-spin interactions \cite{bennett2013phonon,kuzyk2018scaling,maroulakos2020local} and carbon nanotube-mediated coupling \cite{li2016hybrid,dong2019robust,wang2020generation}, a scalable fault-tolerant quantum network is promising. Moreover, the NHQC+ protocols used here for the pure geometric implementation of universal gates should also be of interest to other physical systems such as superconducting circuit, trapped ions, quantum dots, and nuclear magnetic resonance, etc.

We thank Bao-Jie Liu, Man-Hong Yung, and Zheng-Yuan Xue for valuable discussions. This work is supported by the National Key Research and Development Program of China (Grant No. 2017YFA0304504), the National Natural Science Foundation of China (Grant No. 91850102), the Anhui Initiative in Quantum Information Technologies (Grant No. AHY130000), the Science Challenge Project (Grant No. TZ2018003), and the Fundamental Research Funds for the Central Universities (Grant No. WK2030000020).


\clearpage
\newpage
\leftline{\textbf{Supplemental Material}}
\section{NV center in diamond}

The NV center in diamond, a substitution nitrogen atom next to a vacancy, forms a spin triplet system in its ground state. Conveniently, the spin state of NV can be initialized using optical spin pumping under green laser excitation and optically read out by its spin-dependent fluorescence with a home-built confocal microscopy at room-temperature. The NV's electron spin sub-levels of ground state are $\left| 0 \right\rangle $ and $\left| { \pm 1} \right\rangle $, where $\left| {{m_s}} \right\rangle $ denotes the eigenstates of the spin operator ${S_z}$ along the NV's symmetry
axis $z$. The zero-field splitting between $\left| 0 \right\rangle $ and degenerate $\left| { \pm 1} \right\rangle $ sub-levels is ${\text{D} = 2.87}$ GHz. In
the experiment, we apply a strength of the external magnetic field to be near the excited
state level anti-crossing \cite{dong2018non,li2018enhancing}, i.e., ${{B}_0} =378$ G along the NV symmetry axis to split the energy
levels and polarize intrinsic $^{14}{N}$ nuclear spin to enhance the signal contrast up to ${C = 0.25}$.

We employ a room-temperature confocal microscopy, with a dry objective lens (N.A. $= 0.95$ Olympus), to image, initialize and measure NV center in a single-crystal synthetic diamond sample. The NV centers studied in this work are formed during chemical vapor deposition growth \cite{dong2020quantifying}. The abundance of $^{13}C$ is at the nature level of 1\%. Single NV centers are identified by observing anti-bunching in photon correlation measurements and measuring the spin splitting at zero magnetic field. The NV center is mounted on a three-axis closed-loop piezoelectric stage for sub-nanometre-resolution scanning. Fluorescence photons (wavelength ranging from $647$
nm to $800$ nm) are collected into a fiber and detected by the single-photon counting module, with a counting rate of
$50$ kHz and a signal-to-noise ratio of $50:1$ due to a few micrometers below diamond surface. A copper wire of $20$ $\mu m$ diameter above on the bulk diamond is used for the delivery of MW and RF to the NV center. The driving MW is generated by an AWG (Keysight M8190a) and amplified by an MW amplifier (Mini-circuits ZHL-16W-43+). The RF is generated by another AWG (Keysight 33522B) and amplified by another amplifier (Mini-circuits ZHL-32A+). The optical, MW and RF pulse sequences are synchronized by a multichannel pulse generator (Spincore, PBESR-PRO-500).

\section{Signal Interpretation}
In the experiment, the observed fluorescence signal, which is related to the population distributions between ${m_s} = 0 $ and ${m_s} =  \pm 1$ states of NV center, can be converted to the successful probability of quantum coherent operation by linear transformation. Specially, we initialize the system into ${m_s} = 0 $ state by a laser pulse and then change it into ${m_s} =  \pm 1$ state by MW operation and measure their photon counts (denoted by $I_{\max}$, $I_{\min}$). In order to beat the fluctuation of photon counting, we repeat the experimental cycle at least ${10^7}$ times. So the relative population of the ${m_s} = 0 $ state for an unknown state can be expressed as
\begin{equation}
{P_0} = \frac{{I - {I_{\min }}}}{{{I_{\max }} - {I_{\min }}}} \text{,}
 \label{SEq1}
\end{equation}%
where $I$ is the measured photon count under same experimental condition. Therefore, the quantum state of NV center electron spin can be determined from the fluorescence intensity of NV center \cite{dong2018non,dong2020quantifying}.

\section{Optimal Control Sequence}
For resonance NHQC+ scheme \cite{ai2020experimental,li2020fast}, only two time-dependent parameters need to be controlled precisely to enhance the robustness of quantum gates against control amplitude fluctuation. Supposing there is an error ($\alpha $) in control amplitude, the Hamiltonian of the system can be expressed as
\begin{equation}
H_1=(1+\alpha) \frac{\Omega(t)}{2} e^{-i \phi_{1}(t)}|b\rangle\langle a|+\mathrm{H.c.}\text{.}
\end{equation}

\begin{figure*}[tbp]
\centering
\textsf{\includegraphics[width=14cm]{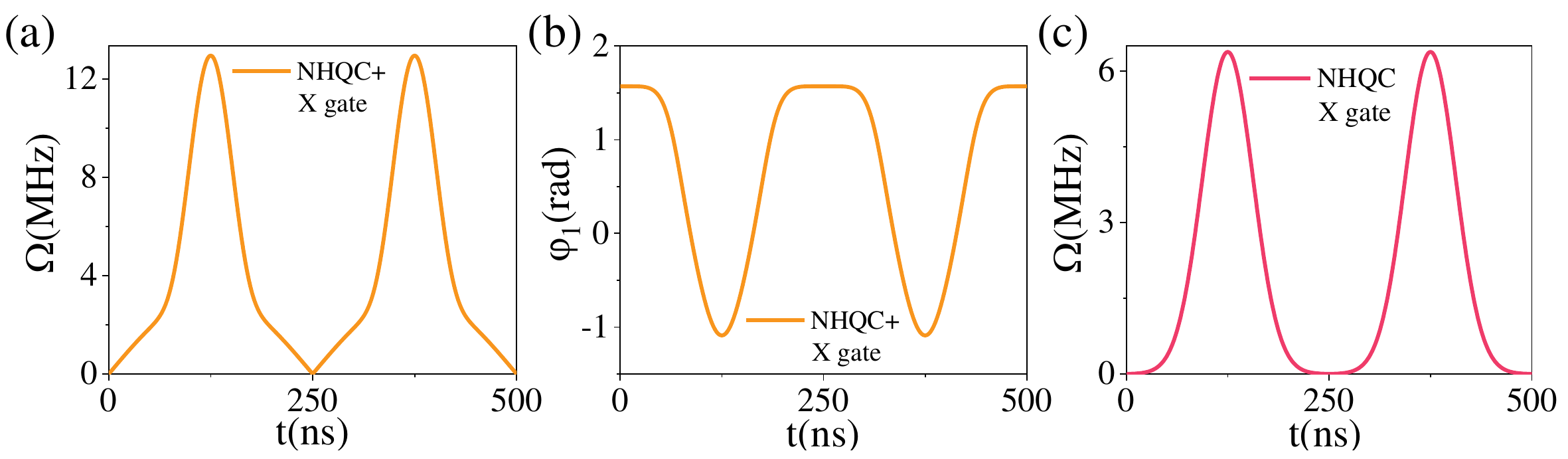}}
\caption{(a)-(b) The amplitude and phase of MW field used in the experiments for the NHQC+ $X$ gate. (c) The waveform pulse of 4$\sigma $ truncated Gaussian shape in conventional NHQC $X$ gate without phase modulation \cite{xu2018single}.}
\label{Sfig1}
\end{figure*}

In our experiment, the influence of the control amplitude errors can be evaluated at the end of the first interval $\tau {\text{/2}}$, and the probability of amplitude errors is given as
\begin{equation}
P=\left|\left\langle\psi(\tau / 2) \mid \psi_{\alpha}(\tau / 2)\right\rangle\right|^{2}=1+O_{1}+O_{2}+\cdots\text{,}
\end{equation}
where $\left| {{\psi _\alpha}} \right\rangle $ is the state with the systematic control error, and
${O_m}$ is the $m$-th order perturbation term. Here, we only calculate the probability amplitude to the second order as
\begin{equation}
P\simeq 1-\alpha^{2}\left|\int_{0}^{\frac{T}{2}} e^{-i f} \dot{\beta } \sin ^{2} \beta d t\right|^{2}\text{.}
\end{equation}
To nullify the error effect \cite{ai2020experimental,li2020fast}, we set $f=\eta (2 \beta-\sin (2 \beta))$, $\varphi(t=0)=0,$ and $\varphi_{2}(t=\tau / 2)=\gamma_G,$ which lead
to $P=1-\sin ^{2}\eta \pi /(2\eta )^{2}.$ In the experimental demonstration, we select $\eta=1$ to show the error-resilience of NHQC+ scheme and waveforms, which are shown in Fig. \ref{Sfig1}(a)-(c).

\begin{figure*}[tbp]
\centering
\textsf{\includegraphics[width=14cm]{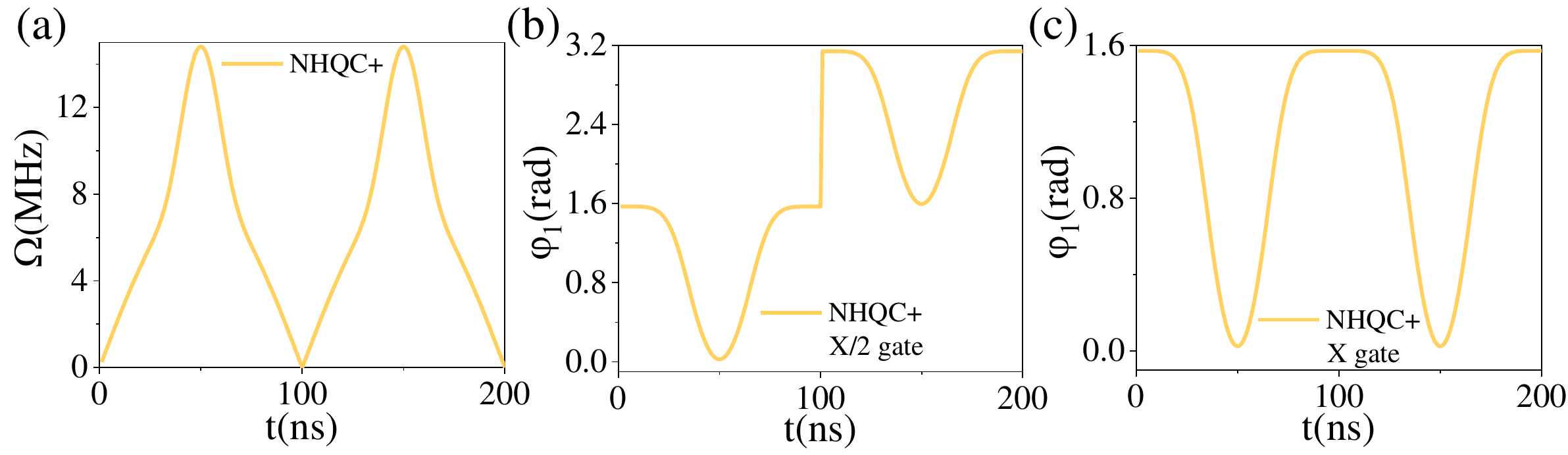}}
\caption{(a) The amplitude of MW field in the experiments for arbitrary holonomic single-qubit gates. (b)-(c) The phases of MW field used in the
experiments for the NHQC+ $X/2$ and $X$ gates with $\eta {\text{ = }}0.4$. For the $Y/2$ gate, $\phi  = \pi /2$ and the amplitude and phase modulation sequence is same as the $X/2$ gate.}
\label{Sfig2}
\end{figure*}

It seems that the optimized pulse is bounded by ${\Omega _{\max }}$, and thus the improvement of the gate performance can only be attributed to the optimal control \cite{li2020fast}. But, in the case of $\eta  \geq 1$, under the restriction, the gate duration time $\tau$ will be too long, and decoherence will introduce unacceptable gate infidelity. Hence, we need to confirm the optimal value of $\eta$ under the targets with both short time $\tau$ and low systematic error sensitivity. In the experiment, we find out that the robustness of the holonomic quantum gates in our scheme is significant improved with $\eta=0.4$ and control sequences are shown in Fig. \ref{Sfig2}(a)-(c).

\section{Quantum process tomography}
To analyze the performance of the NHQC+ scheme, we perform quantum process tomography (QPT) and calculate the fidelity of experimental process \cite{nagata2018universal}. We aim at characterizing the effect of a process $\varepsilon$ on an arbitrary input quantum state $\rho $. Any such processes can be characterized in terms of a dynamical matrix $\chi $, via
\begin{equation}
 \varepsilon (\rho ) = \sum\limits_{i,j = 1}^{\text{4}} {{\chi _{i,j}}} {A_i}\rho A_j^\dag \text{,}
  \label{SEq2}
\end{equation}
where operators ${A_i}$ form a complete basis set of operators.

For a logic qubit system, the tomography matrix $\chi $ has $16$ elements. Four elements follow from the completeness relation $\sum\limits_{i,j = 1}^{\text{4}} {{\chi _{i,j}}} {A_i}A_j^\dag  = E$, and twelve elements have to be measured. By choosing ${A_i}{= \{ I}$, ${\sigma _x}$, ${\sigma _y}$, ${\sigma _z}{\text{\} }}$, where $I$ is the identity and $\sigma_i$ $(i=x$, $y$, $z)$ are the Pauli spin operators, the task reduces to a rather simple procedure. The logic qubit is consecutively prepared in each of the four initial states $\left| \psi  \right\rangle  = \left\{ {\left| 0 \right\rangle ,\left| 1 \right\rangle ,\left( {\left| 0 \right\rangle {\text{ + }}\left| 1 \right\rangle } \right)/\sqrt 2 ,\left( {\left| 0 \right\rangle  - i\left| 1 \right\rangle } \right)/\sqrt 2 } \right\}$. For each initial state, the system evolves driven by the control pulse. To determine the final states $\varepsilon \left( {\left| \psi  \right\rangle \left\langle \psi  \right|} \right)$ by quantum state tomography, we measure the expectation values of the three spin projections ${\sigma _x}$, ${\sigma _y}$, and ${\sigma _z}$ by optical readout \cite{dong2018non}. The readout of the projection on each of the basis states is averaged over ${\text{1}}{{\text{0}}^7}$ shots in experiment. Finally, the complete set of initial states and final states define the tomography matrix $\chi $ by Eq. (\ref{SEq2}). Effects such as noise and finite sampling of the expectation values can lead to one or more negative eigenvalues of the tomography matrix $\chi $ despite the physically required
property of positivity. And by employing maximum likelihood estimation algorithm \cite{nagata2018universal}, this imperfection can be removed effectively.

\end{document}